\documentclass[twocolumn,aps,prb,showpacs,amsmath,amssymb, floatfix]{revtex4}
\usepackage[dvips]{graphicx}
\usepackage{dcolumn}
\usepackage{bm}
\usepackage{color}
\begin{document}
\title{Interacting Systems for Self-Correcting Low Power Switching}
\author{Sayeef Salahuddin}
\email{ssalahud@purdue.edu}
\author{Supriyo Datta}
\affiliation{School of Electrical and Computer Engineering  and NSF Network for Computational Nanotechnology (NCN), Purdue University, West Lafayette, IN-47907, USA.}
\date{\today}
\begin{abstract}
In this paper we first show that dynamic switching schemes can be used to reduce energy dissipation below the thermodynamic minimum of $NkTlnr$ ($N$= number of state variables, $1/r$=error probability), but only at the expense of the error immunity inherent in thermodynamic processes for which the final state is insensitive to the switching dynamics.  It is further shown that, for a system which has internal feedback, e.g. nanomagnets, such that all $N$ spins act in concert, it should be possible to switch with an energy dissipation of the order of $kTlnr$ (considerably less than the thermodynamic limit of $NkTlnr$), while retaining an error immunity comparable to thermodynamic switching. 
\end{abstract}
%
\maketitle
Introduction:
It is generally recognized \cite{labounty:Ref,kish:ref} that the most important factor limiting the down-scaling \cite{borkar:ref} of CMOS devices is the power dissipation, which in the simplest approximation can be described by the charging and discharging of a capacitor. The energy dissipated in charging a capacitor, C, is independent of the wire resistance,R, and is given by
\begin{equation}\label{RC_dissipation}
E_{\text{dissipated}}=lim_{R\rightarrow0}\int^{\inf}_0i^2Rdt=\frac{1}{2}CV^2,
\end{equation}
where V is the applied voltage. The discharging process involves charging the next stage, making the total dissipation in one cycle equal to $CV^2=NqV$, where N is the number of electrons and q is the electronic charge. It can be shown that for an error probability of $1/r=I_{off}/I_{on}$,  thermodynamics requires the minimum voltage to be $V=(kT/q)lnr$, which translates to a theoretical minimum dissipation of $NkTlnr$ or $NkTln2$ (see \cite{cavin:ref, zhirnov:ref} and references therein) for an error probability $1/r=50\%$. 

There is great interest at this time in the possibility of low-power switching through spin-based systems \cite{dmitri:ref,bando1:ref,bando2:ref,cavin:ref}. However, a simple scheme employing a z directed magnetic field to switch  `up' (+z) spins to `down' (-z) will also require a minimum dissipation of $NkTlnr$. To see this, we note that the magnetic field, $B$, creates an energy difference given by $(g\mu_B)B$ between the up and down states so that the error probability, $1/r=N_\uparrow/N_\downarrow=exp(-g\mu_BB/kT)$, where $N_\uparrow$ and $N_\downarrow$ are the occupation probabilities of up-spins and down spins respectively. This requires a minimum $B=kT/(g\mu_B)lnr$ and a minimum energy of $g\mu_BB=kTlnr$ has to be dissipated for each individual spin. Charge and spin based switching thus appear to be very similar with $V\rightarrow B$ and $q\rightarrow g\mu_B$, both of which dissipate $NkTlnr$ for every switching event, involving $N$ entities (charge or spin). 

In this paper, we first show that, with either charge or spin, it is  possible to reduce the dissipation below the limit, $NkTlnr$, by using a system which has an oscillatory response (like an RLC circuit, see Fig. 1), but only at the expense of errors. Next we show that, using interacting systems (like interacting spins in a nanomagnet), it is possible to perform error-free, pseudo-digital switching, while still dissipating considerably less than $NkTlnr$. Using the Landau-Lifshitz-Gilbert (LLG) equation to model realistic nanomagnets, we shall show that a Co cluster of $10^4$ spins can be switched with a power dissipation of only a few $kT$, far less than the thermodynamic limit of $\sim10^4kTlnr$. The basic idea is that spins in a magnet act in concert as a single giant spin, making the dissipation of the order of $kTlnr$ rather than $NkTlnr$. Similar reduction may also be possible using charge-based interacting systems such as ferroelectrics. Note that our approach is different from adiabatic or reversible schemes\cite{landauer:ref} that have been discussed extensively. Although, the total dissipation will be dependent on specific architecture used for communication between computing units, for this paper, we shall restrict ourselves only to the discussion of energy dissipation during an individual switching process.

\emph{Thermodynamic vs. Dynamic logic.-} Using a z directed magnetic field to switch spins from up to down, as discussed in the second paragraph, is an example of what we shall call `thermodynamic switching' where the final state is completely determined by the laws of thermodynamic equilibrium. A different approach is to use a magnetic field perpendicular to the spin direction, say, along the y-axis, wait for the spin to precess exactly by $180^\circ$ and then turn off the field. The precession can be modeled by $d\bar{m}/dt=-\gamma\left(\bar{m}\times\bar{H}\right)$ where, $\bar{m}$  is the polarization of the spin, $\gamma$ is the gyromagnetic ratio and $H$ is the magnetic field. We shall refer to this as `dynamic switching'. Ideally such dynamic switching is completely reversible and no energy ($E=-\bar{m}\cdot\bar{H}$) is dissipated. The price we pay is the extreme sensitivity of the final state to the initial conditions and the duration of the pulse. If the applied pulse is not an exact $\pi$-pulse, an error is incurred which accumulates from one switching event to the next as in an analog computer. To reset the state of the spin reliably, one needs to use thermodynamic switching and pay the energy cost of $NkTlnr$. 

In practice, even dynamic switching will involve some dissipation due to coupling to the environment. The precessional dynamics is described by
\begin{equation}
\label{bloch}
\frac{d\bar{m}}{dt}=-\gamma \bar{m}\times\bar{H}+[T_d]^{-1}\left(\bar{m}-\bar{m_0}\right)
\end{equation}
where, $T_d$ is the damping matrix element and $m_0$ is the initial polarization. A simple calculation shows that the average dissipation will be $E_{\text{dissipated}}\sim(g\mu_BB)/Q_f$, where $Q_f\equiv T_d/\tau_cycle$, is the ratio of the damping time ($T_d$) to the pulse width or half the time period ($\tau_{cycle}$), similar to the quality factor of an RLC circuit. Indeed it can be shown that if we were to charge a capacitor through a series RLC circuit and stop the pulse exactly at point A (see Fig. 1),  then it would be possible to charge the capacitor to supply voltage $V$, while dissipating only $\sim (1/2CV^2)/(Q_f)$ where $Q_f$ denotes the quality factor) \footnote{Interestingly, for dynamic switching $g\mu_BB$ need not equal $kTlnr$ and is only determined by the speed of switching \cite{dmitri:ref}.}. In either case, charge or spin, dissipation is reduced by the `quality factor' but extreme precision is required.
%
\begin{figure}[t]
	\centering
	\includegraphics[width=5 cm]{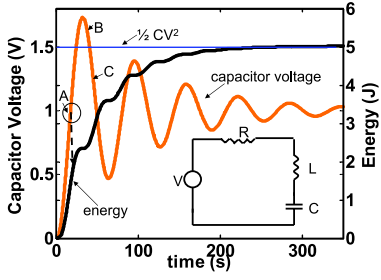}
  \caption{Inset:RLC circuit. The variation of capacitor voltage and dissipated energy with time. The dissipated energy builds up slowly and reaches the value of $1/2CV^2$ as the circuit reaches the steady state with the capacitor charging up to the supply voltage.}
	\label{FIG2}
\end{figure}
The magnetic pulse must be stopped exactly when the spin has rotated by $180^\circ$  and not, for example, by $185^\circ$. Similarly, the RLC circuit needs to be stopped right at point A and not at point B or C. We can correct this error by resetting, only if we immediately dissipate the energy required by thermodynamic processes: for the RLC circuit, we have to let it reach steady state by dissipating  $1/2CV^2$ (see Fig. 1); for spins, we have to put a magnetic field and dissipate $g\mu_BB$, both of which amount to $NkTlnr$ as discussed earlier.

\emph{Self-correcting dynamic logic.-} We now consider a magnet, which is an interacting system of $N$ spins and show that it has the ability to self-correct while dissipating energy $\sim kTlnr$ , far less than the $NkTlnr$ required to switch N non-interacting spins. The magnetization dynamics is described by the Landau-Lifshitz-Gilbert Equation which is written as
\begin{equation}
\left({1+\alpha^2}\right)\frac{\partial \vec{m}}{\partial t}=\gamma\left(\vec{m}\times\vec{H_{eff}}\right)
-\frac{\gamma\alpha}{m}\vec{m}\times\vec{m}\times\vec{H_{{eff}}}
\label{llg}
\end{equation}
where,$H_{\text{eff}}=-\frac{1}{M_s}\nabla_mE$ and $E=-M_s\hat{m}\cdot\bar{H}-K_1cos^2\theta+K_psin^2\theta cos^2\phi$, where $K_1$ is uniaxial anisotropy constant,  $K_p$ is in-plane anisotropy constant, $M_s$ is the saturation magnetization  and  $\alpha$ is the Gilbert damping parameter. The energy landscape of a magnet, with uni-axial anisotropy along the z-axis and easy-plane anisotropy in the y-z plane, (see Fig. 2(a)) is shown in Fig. 2(b). No matter what the initial angle of magnetization, it will relax to one of the two minima at $0$ or $180$, giving rise to the property of self-correction. The energy dissipation in the switching process can be calculated from
\begin{equation}
\frac{dE}{dt}=-\frac{\alpha}{1+\alpha^2}\left(\gamma M_s\right)|\bar{m}\times\bar{H_{\text{eff}}}|^2-M_s\bar{m}\cdot\frac{d}{dt}H_{\text{applied}}
\label{energy_rate}
\end{equation}
\begin{figure}[t]
	\centering
	\includegraphics[width=8cm]{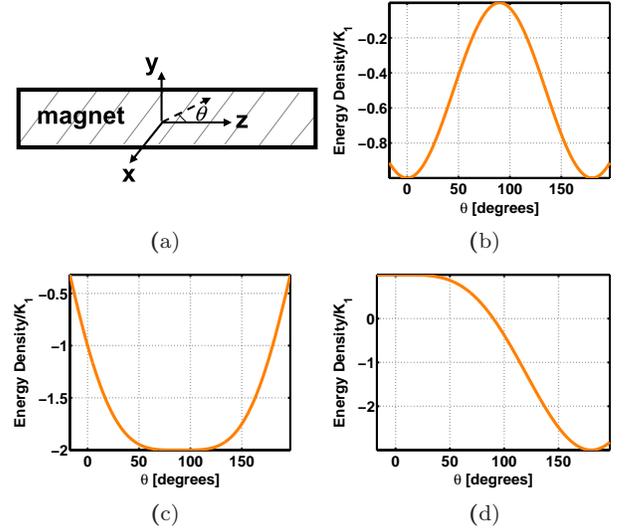}
  \caption{(a) A magnet having an uni-axial anisotropy along the z axis and easy-plane anisotropy in the y-z plane. $\theta$ measures the deflection from the z-axis. (b) Energy landscape without any external field (c) Energy landscape with an applied field in y direction (d) Energy landscape with an applied field in -z direction. For both (c) and (d), it is assumed that magnet always remains in the y-z plane. }
	\label{FIG3}
\end{figure}
Eq. (\ref{energy_rate}) has to be integrated to give the total energy dissipated over the time duration of an applied pulse. In Fig. 3 the solid and dashed curves show the variation of dissipated energy with Gilbert damping coefficient $\alpha$ for a HCP Co with $K_1=3.9\times10^6$ erg/cm$^3$, $K_p=8\pi M_s^2$, $M_s=1400$ emu and volume $\upsilon=70$nm$^3$ (so that in Fig. 2(a) the barrier height, $U=K_1\upsilon=6.6$ kT) for a magnetic pulse applied in y-direction and -z direction respectively. For a typical value of $\alpha=0.1$, the dissipation for each switching event for a y-pulse is $\sim5~kT$ and that for a -z pulse is $\sim35~kT$. Note that the number of spins involved is $(M_s/\mu_B)\upsilon\sim 10^{4}$, for which the  thermodynamic switching would require $10^4kTlnr$. Dynamic switching scheme with self-correction thus requires much smaller dissipation than thermodynamic switching, while retaining the immunity to errors.

\emph{Discussion.-}Let us first look at what happens when a magnetic pulse is applied in the y-direction. The energy landscape is modified from Fig. 2(b) to the form shown in Fig. 2(c) assuming that the magnet always remains in the y-z plane. Instinctively we expect the magnetization to settle in the lowest energy direction, that is the y-direction with $\theta=90$ degrees. But it overshoots and oscillates back and forth several times. If we stop the pulse at any time while $\theta>90^\circ$, the energy landscape will revert to the zero field shape shown in Fig.2(b) and the magnetization will settle to $\theta=180^\circ$. For the parameter values used, the pulse width needs to lie between 20 and 50 ps to achieve this switching. Much longer pulses will take the magnetization to the y-direction with a dissipation of  $E(0)-E(\pi/2)=K_1\upsilon$. But since we rotate the magnet to $\theta=\pi\pm\epsilon$ dynamically and then stop the pulse, the dissipation is lower. 
\begin{figure}[t]
	\centering
	\includegraphics[width=6cm]{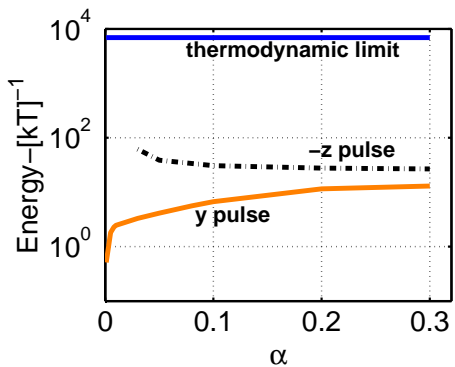}
  \caption{Variation of energy dissipation with Gilbert damping parameter $\alpha$ for y and -z pulses. For comparison, the the thermodynamic limit for dissipation is also plotted.}
	\label{FIG4}
\end{figure}
With a pulse applied in the -z direction (see Fig. 2(d)), there is a lower bound to the pulse width but no upper bound since $\theta=\pi$ is the equilibrium condition (see Fig. 2(d)) for this field direction. However, as we have found (compare $\sim35~kT$  for -z pulse with $\sim5~kT $ for y pulse) this increased error margin comes with the price of higher dissipation. In this case, the minimum dissipation is $E(0)-E(\pi)=4K_1\upsilon$ (see Fig. 2(d)) to which we should add contributions from out of plane motion and $\alpha$.    

It is interesting to note that, switching with a z-pulse, is similar to a thermodynamic process and yet the dissipation ($35~kT$) is much less than the thermodynamic limit ($10^4~kTln2$). This can be understood in the following way. Imagine replacing the system of $N$ particles of charge $q$ each with a single charge $Nq$. Then for an error probability of $1/r$ the minimum voltage would be $V=(kT/Nq)lnr$. Consequently the minimum energy dissipation is $\sim kTlnr$ rather than $NkTlnr$. Similarly, a mono-domain magnet may be thought of a system where all the spins in the volume $\upsilon$ behave as one single spin having a giant magnetic moment of $M_s\upsilon$. In general, for an error probability of $exp(-U/kT)$, a system consisting of $N$ electrons (e.g ferroelectrics) or spins (e.g. ferromagnets), that act together as one, should dissipate overall $\eta U$ (rather than $N\eta U$ for independent particles), where $\eta$ depends on the order of anisotropy(for example, for a second order anisotropy $\eta=4$). This emphasizes the importance of using strongly interacting systems over non-interacting particles \cite{lent1:ref}. The dynamic scheme allows one to further lower the dissipation by making $\eta<1$. Note that this is different from the reversible switching discussed by Landauer and others\cite{landauer:ref,lent1:ref, lent2:ref}.

The barrier height ($U=K_1\upsilon$) determines not only the error probability, but also the retention time, $\tau$, i.e. the time that the magnetic spins would reside in one of the minima shown in Fig. 2(a) before spontaneously transiting to the other. This time can be calculated from $1/\tau=f_0e^{-\beta}$, where $\beta=K_1\upsilon/kT$ and $f_0$ is a constant of the order of $10^9$ s$^{-1}$ (see \cite{aharoni:ref}). In traditional magnets $\upsilon$ is large enough that $\tau$ is days or years and one seldom talks about it. But in our example we have considered a nanomagnet with $\upsilon = 70$ nm$^3$, so that  $K_1\upsilon\sim 6.6~kT$ in order to minimize dissipation. This yields a retention time of $1~\mu$s which should be adequate for logic operations since it represents many clock cycles \cite{cowburn:ref, porod:ref}. If we were to halve the volume, the energy dissipated would be halved, but the retention time would be $\sim30$ ns which may not be acceptable.



In this paper we have not gone into the question of how  the magnetic fields needed for switching are generated. This could either involve traditional `coils' or the recently demonstrated spin-torque effect which could be coupled with magnetic tunneling junction (MTJ) devices to read the information. Since practical spin-torque and MTJ devices operate at a few hundred mV, which is a factor of four to six lower than present-day CMOS supply voltages, low voltage (and hence lower power) operation may be possible provided the resistance of the MTJ is not so high that performance would be compromised. But this does not seem like a fundamental problem and MTJ-spin-torque pairs \cite{salahuddin_apl:ref} with dynamic switching at the device level looks worth investigating. However, we note that unless the MTJ-resistance in the anti parallel combination can be increased signifcantly, MTJs seem more suited to cross bar architectures than conventional CMOS architecture.

\emph{Conclusion.-}The biggest advantage of thermodynamic switching lies in its immunity to errors which comes from the insensitivity of the final state to the details of the switching pulse. But this robust certainty comes with minimum energy cost of $NkTlnr$. We show that we can avoid this energy cost through dynamic switching but the error immunity is lost. We then show, using nanomagnets as an example, that an interacting system, where all $N$ spins act as one, can self-correct and allow an error-free, pseudo-digital switching while dissipating much less than the thermodynamic limit. 
\bibliography{power_dissipation}
\end{document}